\documentclass[%
reprint,
amsmath,
amssymb,
aps,
pra,
superscriptaddress,
preprintnumbers,
prstab,
prstper,groupedaddress]{revtex4-2}
\usepackage{graphicx}
\usepackage{dcolumn}
\usepackage{bm}

\usepackage[english]{babel}
\usepackage[utf8]{inputenc}
\usepackage{amssymb,amsfonts,amsmath,mathtools,mathrsfs}
\usepackage{relsize}
\usepackage{mdframed}
\usepackage{enumitem}
\usepackage{graphicx}

\usepackage[table,usenames,dvipsnames]{xcolor}
\usepackage{gensymb}
\usepackage{grffile}
\usepackage{float}



\usepackage[colorlinks=true,linkcolor=blue,citecolor=blue,urlcolor=blue]{hyperref}

\usepackage{textcomp}
\usepackage{braket} 

\begin{document}

\title{Partial Kondo Screening Solves the Mystery of Rare Earth Tetraborides}

\author{Soumyaranjan Dash}
\affiliation{Department of Physical Sciences, Indian Institute of Science Education and Research (IISER) Mohali, Sector 81, S.A.S. Nagar, Manauli PO 140306, India}

\author{Sanjeev Kumar}
\email{sanjeev@iisermohali.ac.in}
\affiliation{Department of Physical Sciences, Indian Institute of Science Education and Research (IISER) Mohali, Sector 81, S.A.S. Nagar, Manauli PO 140306, India}

\begin{abstract}

We invoke a new mechanism to account for multiple magnetization plateaus observed in rare-earth tetraborides. Using a combination of hybrid and semiclassical Monte Carlo simulations of the Kondo lattice model (KLM) on the Shastry-Sutherland lattice (SSL), we find robust magnetization plateaus at fractions $1/6, 2/9, 1/4, 1/3, 1/2, 2/3, 3/4$ of saturation magnetization. We find that most of the plateau states are partially Kondo screened and  emerge from the field-tuning of a complex three-way competition between the kinetic energy, the Kondo coupling, and the magnetic frustration. Most remarkably, the unusual magnetotransport reported in ErB$_4$ and TmB$_4$ admits an unexpectedly simple explanation within our mechanism. This work not only provides an elegant and simple solution to the long-standing puzzle of metamagnetism and anomalous magnetotransport in RB$_4$ but also introduces a novel mechanism to predict and discover new correlated phases in frustrated Kondo lattices.
\end{abstract}

\date{\today}
\maketitle

{\it Introduction:} Magnetic frustration has long served as a central paradigm for the discovery of new phases, phenomena, and concepts in quantum matter. The insulating frustrated magnets are famous for manifestation of quantum mechanics at macroscale in terms of, for example, quantum spin-liquid phases with exotic excitations \cite{ Anderson1973,Savary2017, Broholm2020, Zhou2017, Balents2010}. In the corresponding metallic systems, such as the frustrated Kondo lattices, the coexistence of itinerant electrons and competing magnetic interactions generates a highly nontrivial interplay that stabilizes unconventional magnetic and electronic phases \cite{Venderbos2012a, Ishizuka2012, Ishizuka2013, Barros2014, Akagi2015, Reja2015a, Reja2016}. The most notable examples are the noncoplanar 4-sublattice state in the triangular lattice \cite{Martin2008,Kumar2010c,Akagi2012} and the realization of the Haldane model on the checkerboard lattice \cite{Venderbos2012a}.

Rare-earth tetraborides, RB$_4$ with R $=$ Tb, Tm, Ho, Nd, Er, are metallic magnets with localized moments residing on the SSL, which belong to the broad class of itinerant frustrated systems. Remarkably, this family exhibits a rich sequence of fractional magnetization plateaus, reminiscent of the quantum Hall effect. For R $=$ Tm, plateaus have been observed at fractional magnetizations $m =1/7,1/8,1/9$, and $1/2$ \cite{Siemensmeyer2008, Michimura2009}, TbB$_4$ displays plateaus with $m=2/9, 1/3, 4/9, 1/2$ and $7/9$ \cite{Yoshii2008, Qureshi2022}, and HoB$_4$ supports a robust plateau at $m=1/3$ along with a narrow region with $m=3/5$ \cite{Brunt2017, Okuyama2008, Kim2009}. 
Early attempts to account for these fractional plateau states (FPS) focused on spin models on the SSL \cite{Moliner2009, Grechnev2013, Chang2009, Meng2008}. However, even with the inclusion of longer-range magnetic interactions, such approaches failed to capture the full range of experimentally observed plateaus \cite{Dublenych2012,Dublenych2013,Wierschem2013, Huo2013}. The problem became more intriguing with the discovery of anomalous electronic transport in TmB$_4$ and ErB$_4$ \cite{Sunku2016, Ye2017}. Both longitudinal and Hall conductivities exhibit strongly non-monotonic field dependence, closely correlated with the metamagnetic response. These observations point to an essential role of conduction electrons in stabilizing FPS in RB$_4$. This motivated studies of itinerancy based mechanisms for magnetism, such as RKKY interactions and double exchange \cite{Feng2014}, however, these approaches did not lead to a satisfactory understanding of the plateau structure. Most importantly, there was no implication for magnetotransport, which is an essential piece of the RB$_4$ puzzle.

In this work, we solve the long-standing mystery of magnetization plateaus and the associated anomalous magnetotransport in RB$_4$ by uncovering a novel partial Kondo screening mechanism. The mechanism is most transparently illustrated in the strong coupling limit of the KLM, corresponding to a variant of the Kondo necklace model (KNM). Using semiclassical Monte Carlo (SMC) simulations of the KNM on the SSL, we show that the competition between Kondo singlet formation and magnetic frustration generates a variety of FPS even in the absence of electron itinerancy. In addition, we  perform hybrid Monte Carlo (HMC) simulations of the KLM to show that the electronic kinetic energy further enriches the plateau structure of magnetization. Importantly, the mechanism invoked in this work provides, for the first time, a natural and unified explanation of the anomalous magnetotransport observed in ErB$_4$ and TmB$_4$.

\begin{figure}
\includegraphics[width=0.62 \columnwidth,angle=0,clip=true]{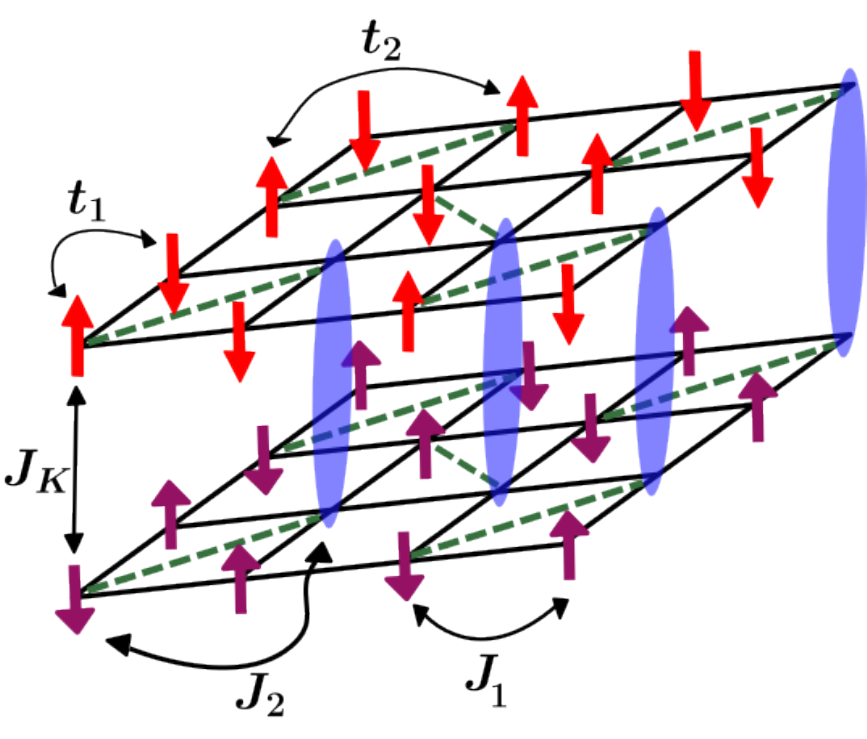}
 \caption{A schematic representation of our semiclassical version of the KLM. Top layer is the tight-binding model on the SSL with $t_1, t_2$ denoting the hopping parameters on SSL. The bottom layer represents localized moments interacting via $J_1$ and $J_2$.  The vertical ovals indicate local singlets at C sites. The U sites retain a Zeeman-like coupling of strength $J_{\rm K}$.}
\label{Fig:M1}   
\end{figure}

{\it The Model and the Semiclassical Approach:}
We consider the KLM on SSL in the presence of an external magnetic field. The Hamiltonian, which has contributions from the electronic kinetic energy, the Kondo coupling between the itinerant and localized spins, and the magnetic exchange interactions, is given by

\begin{equation}
\begin{aligned}
H &= H_{\rm Kin} + H_{\rm Kondo} + H_{\rm Spin}, \\
H_{\rm Kin} &= -t_1 \sum_{\langle ij \rangle,\sigma} ( c_{i\sigma}^\dagger c_{j\sigma} + \mathrm{H.c.} )
- t_2 \sum_{\langle\langle ij \rangle\rangle,\sigma} ( c_{i\sigma}^\dagger c_{j\sigma} + \mathrm{H.c.} ), \\
H_{\rm Kondo} &= J_{\rm K} \sum_i \mathbf{s}_i \cdot \boldsymbol{\sigma}_i, \\
H_{\rm Spin} &= J_1 \sum_{\langle ij \rangle} \mathbf{s}_i \cdot \mathbf{s}_j 
+ J_2 \sum_{\langle\langle ij \rangle\rangle} \mathbf{s}_i \cdot \mathbf{s}_j 
- h_z \sum_i s_i^z,
\end{aligned}
\label{eq:Ham1}
\end{equation}
\noindent 
where $c^{\dagger}_{i,\sigma}$ ($c^{}_{i,\sigma}$) are the creation (annihilation) operators for electrons at the site $i$ with spin $\sigma = \uparrow, \downarrow$, and ${\bf s}_i$ ($\sigma_i$) are spin operators for localized (itinerant) spin-1/2 variables. The hopping parameters $t_1, t_2$ and the exchange constants $J_1, J_2$ are schematically shown in Fig. \ref{Fig:M1}. The reference energy scale is set by $J_1 = 1$ and $h_z$ denotes the strength of the external magnetic field. For a clear illustration of the new mechanism, we consider a minimal single-channel KLM. For R = Er, Tm, Tb an effective spin-$1/2$ description for localized spin is justified due to the presence of strong crystal fields \cite{Siemensmeyer2008, Suzuki2009, Suzuki2010}. The reference energy scale is set by $J_1 = 1$ and we use $J_2/J_1 = 1$ and $t_2/t_1 = 1$, unless stated otherwise.

The Hamiltonian Eq. (\ref{eq:Ham1}) defines a formidable many-body problem, with numerical approaches severely limited by the exponential growth of the Hilbert space. We therefore adopt a semiclassical approximation to render the problem tractable. The key idea is to introduce effective classical variables that represent the formation of a local Kondo singlet \cite{Dash2025}. However, the tendency to form such local Kondo singlets competes with the presence of magnetic order preferred by direct exchange terms $J_1, J_2$ and conduction electron mediated interactions. We retain this complex three-way competition in the simplified semiclassical Hamiltonian defined as

\begin{equation}
\begin{aligned}
H =\;& H_{{\rm Kin}} + J_{\rm K} \sum_{i \in { {\cal U}} } {\cal S}_i {\bf \sigma}^z_i
+ J_{\rm K} \sum_{i \in {{ \cal C}}, \sigma} c_{i\sigma}^\dagger c_{i\sigma}
  + N E_{\rm K} n_{\rm K}\\
& + J_1 \sum_{\langle ij \rangle \in {{\cal U}} } {\cal S}_i ~ {\cal S}_j 
  + J_2 \sum_{\langle \langle ij \rangle \rangle \in {{\cal U}}} {\cal S}_i ~ {\cal S}_j
   - h_z \sum_i {\cal S}_i.
\end{aligned}
\label{eq:Ham2}
\end{equation}
\noindent
In the above, the Kondo coupling term is assigned a dual character. At certain sites, marked as correlated (${\cal C}$), the Kondo coupling leads to the formation of perfect singlets with the associated energy of each singlet given by $E_{{\rm K}} = \frac{-3J_{\rm K}}{4}\left( \dfrac{1-e^{-J_{\rm K}/T}}{1+3e^{-J_{\rm K}/T}} \right)$. On the remaining sites, marked as uncorrelated (${\cal U}$), the Kondo coupling simply provides an effective magnetic field for conduction electrons. This motivates the replacement of quantum operators by classical variables ${\cal S}_i$ at the ${\cal U}$ sites that, for simplicity, are assigned an Ising character. Most importantly, the number and the real-space distribution of the ${\cal U}$ and ${\cal C}$ sites are dynamical and completely controlled by the energetics. $n_{\rm K}$ is the fraction of ${\cal C}$ sites, which also modifies the effective density of conduction electrons by $n_c = n^0_c - n_{\rm K}$, where $n^0_c$ denotes the bare density of conduction electrons. The third term in Eq. (\ref{eq:Ham2}) represents the effective on-site potential  at the ${\cal C}$ sites experienced by conduction electrons not trapped in Kondo singlets \cite{Dash2025}.

Despite the simplifications, the analysis of the semiclassical Hamiltonian Eq. (2) remains a challenging optimization problem due to the dynamical character of sites as ${\cal U}$ or ${\cal C}$, and the spin degrees of freedom at the ${\cal U}$ sites. The hybrid Monte Carlo (HMC) provides a numerically exact approach to study the semiclassical Hamiltonian (see Supplemental Material for details). However, accessible lattice sizes in the HMC remain limited to $N \sim 100$. In the limit where $J_{\rm K}$ and $J_1, J_2$ dominate over the kinetic energy, the competition between the Kondo and the exchange terms is tuned by the external magnetic field. In this limit, the kinetic energy term may be completely neglected, leading to a variant of the KNM. The resulting model, together with the classical description of local singlets, facilitates large-scale SMC simulations (see Supplemental Material for details).

\begin{figure}[t!]
\includegraphics[width=0.96 \columnwidth,angle=0,clip=true]{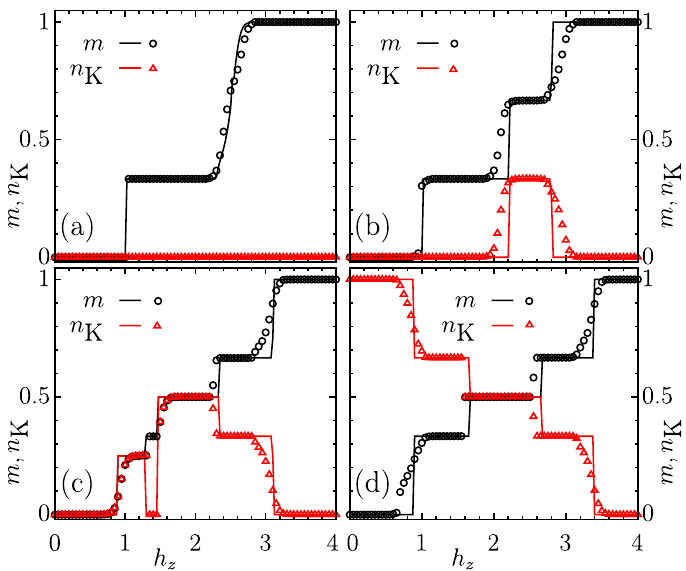}
 \caption{Variations of the normalized magnetization, $m$, and the fraction of Kondo singlets, $n_{\rm K}$, with external field, $h_z$, for (a) $J_{\rm K} = 0.0$, (b) $J_{\rm K} = 0.3$, (c) $J_{\rm K} = 0.6$, and $J_{\rm K} = 0.9$.  The symbols are the SMC data and the continuous lines are obtained from variational calculations (see Supplemental Material for details).}
\label{Fig:M2}   
\end{figure}
\begin{figure*}
\includegraphics[width=1.75 \columnwidth,angle=0,clip=true]{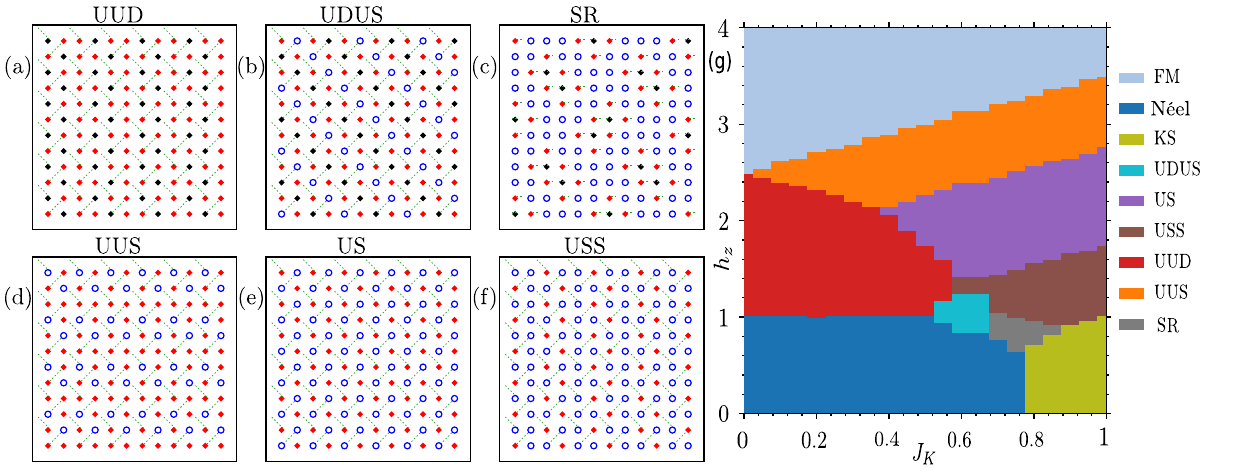}
 \caption{Real-space patterns of spins (red: ${\cal S} = \tfrac{1}{2}$, black: ${\cal S} = -\tfrac{1}{2}$) and singlets (blue circles) of a $12 \times 12$ lattice section corresponding to different FPS identified via the values of $m$ and $n_{\rm K}$ as, (a) UUD: $m= 1/3$, $n_{\rm K} = 0$, (b) UDUS: $m= 1/4$, $n_{\rm K} = 1/4$, (c) SR (singlet rings): $m= 2/9$, $n_{\rm K} = 11/20$, (d) UUS: $m= 2/3$, $n_{\rm K} = 1/3$, (e) US: $m= 1/2$, $n_{\rm K} = 1/2$, and (f) USS: $m= 1/3$, $n_{\rm K} = 2/3$ as obtained from SMC simulations on $24 \times 24$ lattice. (g) Phase diagram in $J_{\rm K} - h_z$ plane.}
\label{Fig:M3}   
\end{figure*}

{\it Magnetization Plateaus in the Kondo necklace limit:}
We begin by studying field-induced magnetization in the $t_1 = t_2 = 0$ limit of the Hamiltonian Eq. (\ref{eq:Ham2}), which corresponds to a variant of the KNM. The semiclassical model consists of two types of degrees of freedom: (i) each site can be of type ${\cal U}$ or ${\cal C}$ and (ii) on the ${\cal U}$ sites the local moment can be $\pm 1/2$. We make use of SMC simulations to study the evolution of magnetization with the external magnetic field (see Supplemental Material for details). In the absence of Kondo coupling, the model reduces to a frustrated Ising model. In this limit, for $J_1 = J_2 = 1$, we find only one non-trivial plateau state with $m = 1/3$ (see Fig. \ref{Fig:M2}(a)) \cite{Chang2009, Meng2008, Dublenych2012}. For $J_{\rm K} = 0.3 $, a new plateau with $n_{\rm K} = 1/3$ appears at $m = 2/3$ (see Fig. \ref{Fig:M2}(b)). Two new FPS are supported at $J_{\rm K} = 0.6$, with multiple non-monotonic changes in the singlet count with external magnetic field (see Fig. \ref{Fig:M2}(c)). Finally, $J_{\rm K} = 0.9$ represents the strong Kondo limit where the zero-field state is the Kondo screened (KS), and the fraction of Kondo singlets decreases monotonically with increasing field (see Fig. \ref{Fig:M2}(d)). In order to determine a comprehensive ground state phase diagram in the $h_z - J_{\rm K}$ plane, we use a variational ansatz where different FPS identified in the simulations are compared in energy. The results of the variational approach compare very well with those obtained in the SMC simulations (Fig. \ref{Fig:M2}(a)-(d)).

The real-space configurations corresponding to all the non-trivial FPS are displayed in Fig. \ref{Fig:M3}(a)-(f).
We label these configurations in terms of the patterns of up (U), down (D) and singlet (S) sites. The six non-trivial FPS obtained in the KNM limit are, UUD (Fig. \ref{Fig:M3} (a)), UDUS (Fig. \ref{Fig:M3} (b)), singlet rings (SR) ((Fig. \ref{Fig:M3} (c)), UUS (Fig. \ref{Fig:M3} (d)), US (Fig. \ref{Fig:M3} (e)) and USS (Fig. \ref{Fig:M3} (f)).
The stability regions of different FPS are summarized in the phase diagram in the $h_z - J_{\rm K}$ plane (see Fig. \ref{Fig:M3}(g)). In the conventional description, the partial magnetization in Ising systems arises from different arrangements of up and down spins only. Our new mechanism allows the possibility of KS sites with vanishing moments, and hence new values of magnetization fractions become stable. We note that a simple classical description of the KNM already accounts for a larger number of FPS compared to the approaches based purely on magnetic exchange interactions. Within our semiclassical description, the emergent pattern of KS sites can be viewed as a realization of site-selective spontaneous symmetry breaking \cite{Dash2025}. In an alternative scenario, quantum fluctuations may restore the translational invariance, resulting in a spatially uniform reduction of moment sizes. Nevertheless, in either case, the fractional value of the magnetization remains unchanged. Neutron scattering experiments on TmB$_4$ suggest the presence of modulated local moments, supporting the possibility of inhomogeneous Kondo screening \cite{Wierschem2015}.

{\it Effect of electron itinerancy:}
Transport measurements confirm that RB$_4$ are good metals, which implies that itinerant electrons play an essential role in their low-energy physics. It is therefore important to verify that the FPS, obtained within the KNM limit, remain robust upon incorporating electron itinerancy. To address this, we study the model for finite values of the hopping amplitudes $t_1$ and $t_2$. An explicit treatment of the kinetic-energy term requires the use of HMC simulations, where the effective one-particle quantum problem is solved numerically at each Monte Carlo step (see Supplemental Material for details). Compared to the SMC simulations used in the KNM limit, the HMC simulations are computationally more demanding, however, the advantage is in terms of an explicit simultaneous treatment of the Kondo screening, the magnetic frustration, and the electron delocalization.

\begin{figure}
\includegraphics[width=0.96 \columnwidth,angle=0,clip=true]{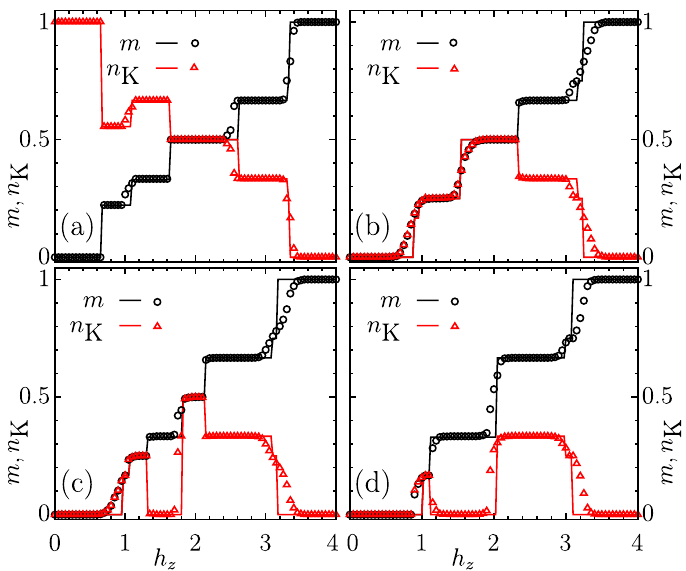}
 \caption{Variations of normalized magnetization, $m$, and fraction of Kondo singlets, $n_K$ with $h_z$, obtained from variational calculations and HMC simulations for $J_K$ = 0.90, and (a) $t_1 = 0.10$, (b) $t_1 = 0.15$, (c) $t_1 = 0.20$, and $t_1 = 0.25$. }
\label{Fig:M4}   
\end{figure}

The kinetic-energy term disfavors the formation of local Kondo singlets, rendering the Kondo-screening mechanism ineffective at small $J_{\rm K}$. Consequently, the new mechanism is expected to remain relevant in the presence of hopping only for sufficiently large $J_{\rm K}$. For weak hopping, the zero-field ground state remains fully Kondo screened. However, the field dependence already exhibits a nonmonotonic behavior of $n_{\rm K}$ at $t_1=0.1$, accompanied by the emergence of a new plateau state absent in the Kondo necklace model [Fig. \ref{Fig:M4}(a)]. Increasing the hopping strength leads to further modifications in the field dependence of both the Kondo-singlet fraction and the magnetization [Figs. \ref{Fig:M4}(b)-(d)]. The resulting multiply  nonmonotonic changes in $n_{\rm K}$ as a function of external magnetic field highlight the nontrivial competition among kinetic energy, Kondo coupling, and magnetic frustration. The inclusion of kinetic energy term stabilizes two new FPS with $m=1/6$ and $m=3/4$. The variations of $n_{\rm K}$ with magnetic field play a crucial role in resolving the puzzle of anomalous magnetotransport response in some of the RB$_4$ compounds, as we will discuss later.

\begin{figure}
\includegraphics[width=0.96 \columnwidth,angle=0,clip=true]{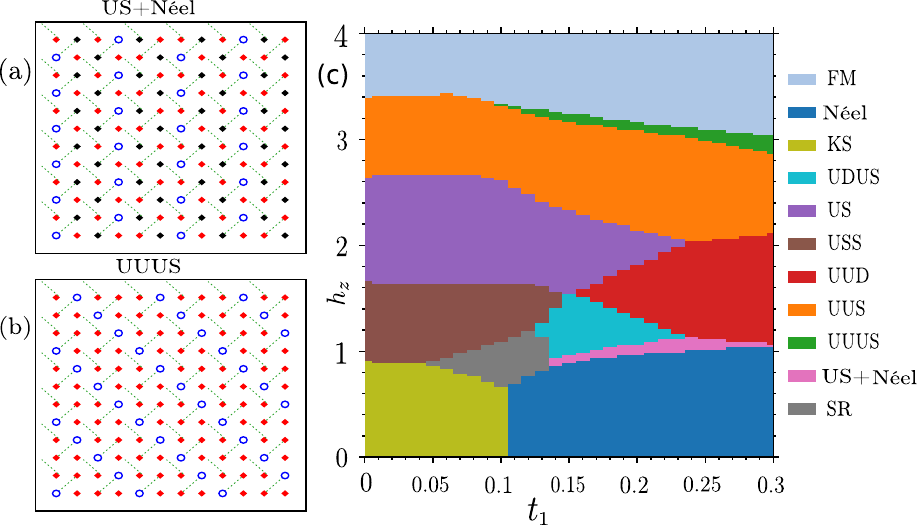}
 \caption{Real-space patterns of spins (red: ${\cal S} = \tfrac{1}{2}$, black: ${\cal S} = -\tfrac{1}{2}$) and singlets (blue circles) of a $12 \times 12$ lattice section corresponding to (a) US+N\'eel: $m= 1/6$, $n_{\rm K} = 1/6$, (b) UUUS: $m= 3/4$, $n_{\rm K} = 1/4$. (c) phase diagram in $t$ - $h_z$ plane for $J_2/J_1 = 1.0$ and $J_{\rm K} = 0.90$}
\label{Fig:M5}   
\end{figure}

We find that all the FPS of the KNM remain stable upon inclusion of kinetic-energy terms. In addition, two new ground states emerge with magnetization fractions $1/6$ and $3/4$, which are shown in Figs. \ref{Fig:M5}(a)–(b). Using variational calculations, we map the stability regions of the different ground states in the $h_z-t_1$ plane. The resulting phase diagram at $J_{\rm K}=0.9$ exhibits eight distinct plateau states at nontrivial fractions of magnetization [Fig. \ref{Fig:M5}(c)]. The mechanism proposed here for the existence of FPS has the potential to describe other fractions as well. Here, all results are presented for the case of half-filling of the electronic single band. Different choices of filling fractions of the effective single band may support other FPS. Furthermore, the number of plateau phases reported here may be limited by the accessible cluster sizes within HMC. 

{\it Explanation of Anomalous Magnetotransport:}
A key unresolved aspect of RB$_4$ is their highly anomalous magnetotransport. In particular, ErB$_4$ and TmB$_4$ exhibit pronounced variations in both longitudinal and Hall conductivities across different magnetic plateau phases \cite{Ye2017, Sunku2016}, which remain unexplained within existing theoretical frameworks \cite{Sunku2016}. We show that the mechanism proposed here naturally accounts for these anomalous magnetotransport signatures within a simple effective Drude description.

Within our semiclassical approach, a fraction $n_{\rm K}$ of conduction electrons get trapped in Kondo singlets leading to a reduction in the effective number of conduction electrons. Complete localization of this type is known to occur only in the strong-coupling limit, as realized in Kondo insulators \cite{Gabani2003}. In metallic systems, where Kondo screening coexists with charge transport, a realistic picture is that of enhancement of the effective mass of charge carriers.
If a fraction $f$ of the carriers acquire a heavy mass $M$, while the remaining fraction $(1-f)$ retains the bare mass $m_0$, the effective mass relevant for transport is given by, $1/m^* = f/M + (1-f)/m_0 $. We use $f = n_{\rm K}$ and $M = \alpha ~ m_0$  with $\alpha$ as the mass enhancement factor due to Kondo screening. The Drude conductivities within this simple approach are given by,

\begin{align}
\sigma_{xx}
&=
\frac{
n^0_c e^2 \tau / m^{*}
}{
1+\left(
\dfrac{e \tau B_{\rm eff}}{m^{*}}
\right)^2
},
\label{eq:sigmaxx}
\\[8pt]
\sigma_{xy}
&=
-
\frac{
n^0_c e^3 \tau^2 B_{\rm eff}/(m^{*})^2
}{
1+\left(
\dfrac{e \tau B_{\rm eff}}{m^{*}}
\right)^2
},
\label{eq:sigmaxy}
\end{align}
\noindent where $n^0_c$ is the bare electron density, $e$ represents the magnitude of electronic charge, $\tau$ is the Drude mean-free time, $m^{*}$ is the effective electron mass and $B_{\rm eff}$ is the effective magnetic field experienced by the electrons.

\begin{figure}
\includegraphics[width=0.96 \columnwidth,angle=0,clip=true]{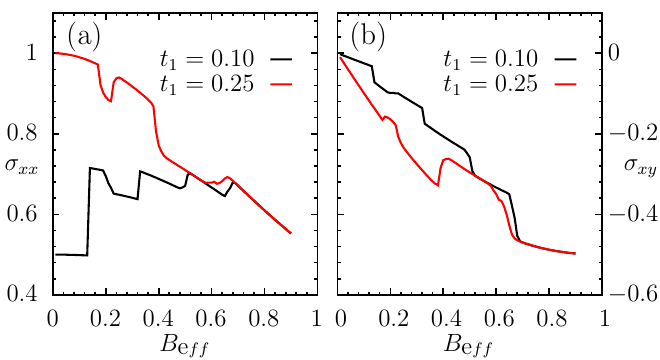}
 \caption{Longitudinal (a) and transverse (b) conductivities as a function of magnetic field within the mass-enhanced Drude approach for two different values of hopping parameter $t_1$. The variations shown here compare very well with the experimental data shown in Figs. 4 and 5 in Ref. \cite{Ye2017} }
\label{Fig:M6}   
\end{figure}
Since our primary focus here is to describe the variations of electronic transport with magnetic field, we set $e = m_0 = \tau = 1$. We use $\alpha = 2$ as the mass enhancement factor. We find that $\sigma_{xx}$ exhibits pronounced anomalies at the magnetic fields corresponding to transitions between successive plateau phases (Fig. \ref{Fig:M6}(a)). We find that in some cases the sign of magnetoresistance can be negative, which compares remarkably well with the data on TmB$_4$ \cite{Ye2017}. Similarly, change in slopes in $\sigma_{xy}$ associated with transitions between successive plateau phases is clearly explained within the effective Drude approach (Fig. \ref{Fig:M6}(b)). In an earlier study, the unusual magnetic field dependence was fitted via a non-trivial field-dependence of $\tau$ without assigning any mechanism to this variation \cite{Ye2017}. Nevertheless, this approach validated the use of Drude model for describing electronic transport in RB$_4$.

{\it Conclusions:}
We resolve the long-standing puzzle of a plethora of fractional magnetization plateaus observed in rare-earth tetraborides by identifying a novel mechanism based on partial Kondo screening. Through explicit simulations of the KLM treated at a semiclassical level, we uncover plateau phases at magnetization fractions 1/6, 2/9, 1/4, 1/3, 1/2, 2/3, and 3/4, many of which have already been reported in experiments on  RB$_4$. No existing theory has accounted for such a diverse set of plateau states within a unified framework. The essential insight is that selective formation of Kondo singlets greatly enlarges the configurational space available to the magnetic moments that survive Kondo screening, thereby stabilizing multiple FPS. Conventionally, strong external magnetic field breaks Kondo singlets, however, the magnetic frustration, the kinetic energy and the Kondo coupling combine in such a way that the external magnetic field favors the formation of Kondo singlets, generating the possibility of unexpected FPS. More importantly, incorporating the effect of partial screening and its evolution with magnetic field into the transport calculations naturally explains the puzzling anomalous magnetotransport data of RB$_4$. 
On the one hand, we establish the relevance of a previously overlooked mechanism for magnetization plateau formation in rare-earth tetraborides. On the other, we uncover a general mechanism for metamagnetism and anomalous magnetotransport that is broadly applicable to geometrically frustrated Kondo systems. Material-specific extensions incorporating multichannel Kondo couplings and larger effective spins constitute promising directions for future work and are expected to reveal an even richer hierarchy of fractional plateau phases.

{\it Acknowledgements :} We acknowledge the use of the HPC facility at IISER Mohali. S. D. acknowledges IISER Mohali for support through the institute fellowship.

\begin{widetext}


\section*{Supplemental Material}

\setcounter{section}{0}
\setcounter{equation}{0}
\setcounter{figure}{0}

\renewcommand\thesection{\Alph{section}}
\renewcommand{\theequation}{S\arabic{equation}}
\renewcommand{\thefigure}{S\arabic{figure}}

\renewcommand{\theHfigure}{S\arabic{figure}}
\renewcommand{\theHequation}{S\arabic{equation}}

\section{Semiclassical Monte Carlo Simulations}

The Hamiltonian Eq.~(\ref{eq:Ham2}) in the main text admits a simple semiclassical treatment in the limit \(t_1=t_2=0\). Since the kinetic energy term vanishes, the Hamiltonian reduces to a purely local problem governed entirely by the interaction terms. This can be efficiently treated via a semiclassical Monte Carlo (SMC) method, where the spin and charge variables are effectively classical.
The equilibrium properties of the model are obtained by sampling these classical configurations using standard Metropolis updates, allowing access to a much larger lattice size.

Each lattice site is characterized by two local degrees of freedom. The first is a binary variable that specifies whether the site is in the correlated (\({\cal C}\)) or uncorrelated (\({\cal U}\)) local configuration, as defined in the main text. The second is a local Ising variable, \({\cal S}_i=\pm \tfrac{1}{2}\), which is defined only on the \({\cal U}\) sites. The \({\cal C}\) sites are nonmagnetic at $T=0$ and do not carry a local moment. Physically, the \({\cal U}\) sites correspond to regions where the Kondo coupling acts as an effective magnetic field on the conduction electrons, motivating the replacement of the corresponding quantum degrees of freedom by classical Ising variables. In contrast, the \({\cal C}\) sites represent locally correlated regions without the local magnetic moments have been screened by the conduction cloud. Importantly, both the number of \({\cal C}\) and \({\cal U}\) sites, as well as their real-space arrangement, are treated as dynamical variables determined entirely by the energetics of the system. The energy of a given configuration is therefore governed by the local energy competition between the \({\cal C}\) and \({\cal U}\) sites, together with the Zeeman coupling of the \({\cal U}\)-site moments to the external magnetic field.

Configurations are sampled using the standard Metropolis algorithm. During each Monte Carlo step, a lattice site is selected randomly and one of two possible updates is proposed: (i) a change of the local state between the uncorrelated (\({\cal U}\)) and correlated (\({\cal C}\)) configurations, or (ii) a flip of the local Ising moment, \({\cal S}_i \rightarrow -{\cal S}_i\), when the selected site belongs to the \({\cal U}\) sector. For \({\cal U}\leftrightarrow{\cal C}\) updates, a random spin orientation is assigned when a site changes from \({\cal C}\) to \({\cal U}\), while the local moment is removed when a site changes from \({\cal U}\) to \({\cal C}\). Since the quantum electronic degrees of freedom are replaced by classical local variables while retaining the energetics associated with the underlying correlated and uncorrelated sites, the resulting approach constitutes a semiclassical Monte Carlo scheme.

For most model parameters, the simulations are performed on a \(24\times24\) lattice with periodic boundary conditions. For selected parameters, the stability of the results is ensured by simulating lattice sizes as large as \(60\times60\). The system is initialized from random configurations at high temperature and annealed to the desired low temperature using a sufficient number of temperature steps. At each temperature, the system is equilibrated over $10^5$ Monte Carlo steps (MCS), followed by an additional $10^5$ steps for computation of relevant quantities. Once the target temperature is reached, an external magnetic field is applied and varied systematically. For each value of the field, the system is equilibrated over $10^5$ MCS, followed by another $10^5$ MCS for computations. To ensure convergence, the magnetic field is varied incrementally, using the final configuration at a given field as the initial condition for the next field value.

The fractional magnetization is calculated from the average of the localized moments residing on the \({\cal U}\)-type sites and is defined as \(m = \frac{2}{N}\sum\limits_{i \in {\cal U}} {\cal S}_i\), while the fraction of \({\cal C}\)-type sites is given by \(n_{\rm K} = \frac{1}{N}\sum\limits_{i \in {\cal C}} 1\), where \(N\) is the total number of lattice sites. This SMC approach provides a controlled description of the field-induced magnetization in the absence of fermionic dynamics and serves as a reference for the results obtained by the hybrid Monte Carlo approach, which is discussed in the next section.

\section{Hybrid Monte Carlo Simulations}

The effective Hamiltonian introduced in Eq.~(\ref{eq:Ham2}) of the main text contains both classical and quantum degrees of freedom. The classical degrees of freedom consist of the real-space arrangement of correlated (\({\cal C}\)) and uncorrelated (\({\cal U}\)) lattice sites, together with the localized Ising moments \({\cal S}_i\) defined on the \({\cal U}\) sites. The itinerant electrons are described by the fermionic creation and annihilation operators, \(c_i^\dagger\) and \(c_i\). Within the semiclassical approximation, the localized moments are treated as classical variables, rendering the fermionic sector quadratic. Consequently, for a given real-space configuration of \({\cal C}\)- and \({\cal U}\)-type sites together with the localized moments \({\cal S}_i\), the many-body problem reduces to a quadratic single-particle Hamiltonian that can be diagonalized exactly, with the electronic contribution obtained from the corresponding single-particle eigenstates. The dimension of the single-particle Hilbert space therefore scales linearly with the system size.

The ground-state configuration is determined by minimizing the total energy, including both the classical and fermionic contributions. To achieve this, we employ a hybrid Monte Carlo approach in which the classical variables are updated using the Metropolis algorithm, while the fermionic contribution to the energy is evaluated through exact diagonalization of the corresponding single-particle Hamiltonian for every proposed configuration. During each Monte Carlo update, a lattice site is selected randomly and one of the following updates is proposed: (i) a change of the selected site between the \({\cal C}\) and \({\cal U}\) types, or (ii) a flip of the local Ising moment, \({\cal S}_i \rightarrow -{\cal S}_i\), if the site belongs to the \({\cal U}\) sector. For \({\cal C}\leftrightarrow{\cal U}\) updates, a random spin orientation is assigned when a site changes from \({\cal C}\) to \({\cal U}\), while the local moment is removed when a site changes from \({\cal U}\) to \({\cal C}\). Since the Hamiltonian is diagonalized at every Monte Carlo update, the computational cost scales as \(N^4\), where \(N\) is the number of lattice sites.

The simulations are performed on two-dimensional \(6\times6\), \(6\times8\), and \(8\times8\) lattices with periodic boundary conditions. Here, all results are presented for the
case of half-filling of the electronic single band and $t_2/t_1$ ratio is kept fixed as 1. The ground state is identified by comparing the total energies obtained for different lattice sizes. The fermionic Hamiltonian is diagonalized using the \texttt{CHEEVX} routine from the LAPACK library. For each parameter set, \(10^5\) Monte Carlo steps are used for equilibration, followed by an additional \(10^5\) steps for measurements. The simulations are initialized at high temperature using random configurations of the \({\cal C}\)- and \({\cal U}\)-type sites together with random \({\cal S}_i\) configurations, and the system is gradually cooled in a manner similar to the semiclassical Monte Carlo simulations.

\section{Variational Calculations}

To complement the SMC and HMC results and to reduce finite-size effects, we perform variational calculations on significantly larger lattices. These calculations are designed to provide a more reliable identification of the ground-state phases and phase boundaries. The set of candidate states considered in the variational approach is guided by the configurations observed in the hybrid Monte Carlo simulations. In practice, we construct representative spin and singlet configurations corresponding to the distinct phases identified numerically and evaluate their energies independently. All variational calculations are carried out on a two-dimensional square lattice of size $120 \times 120$ with periodic boundary conditions. The ground state is then determined by selecting the configuration with the lowest total energy among all candidates.

\begin{figure}[h!]
\includegraphics[width=0.95\columnwidth,angle=0,clip=true]{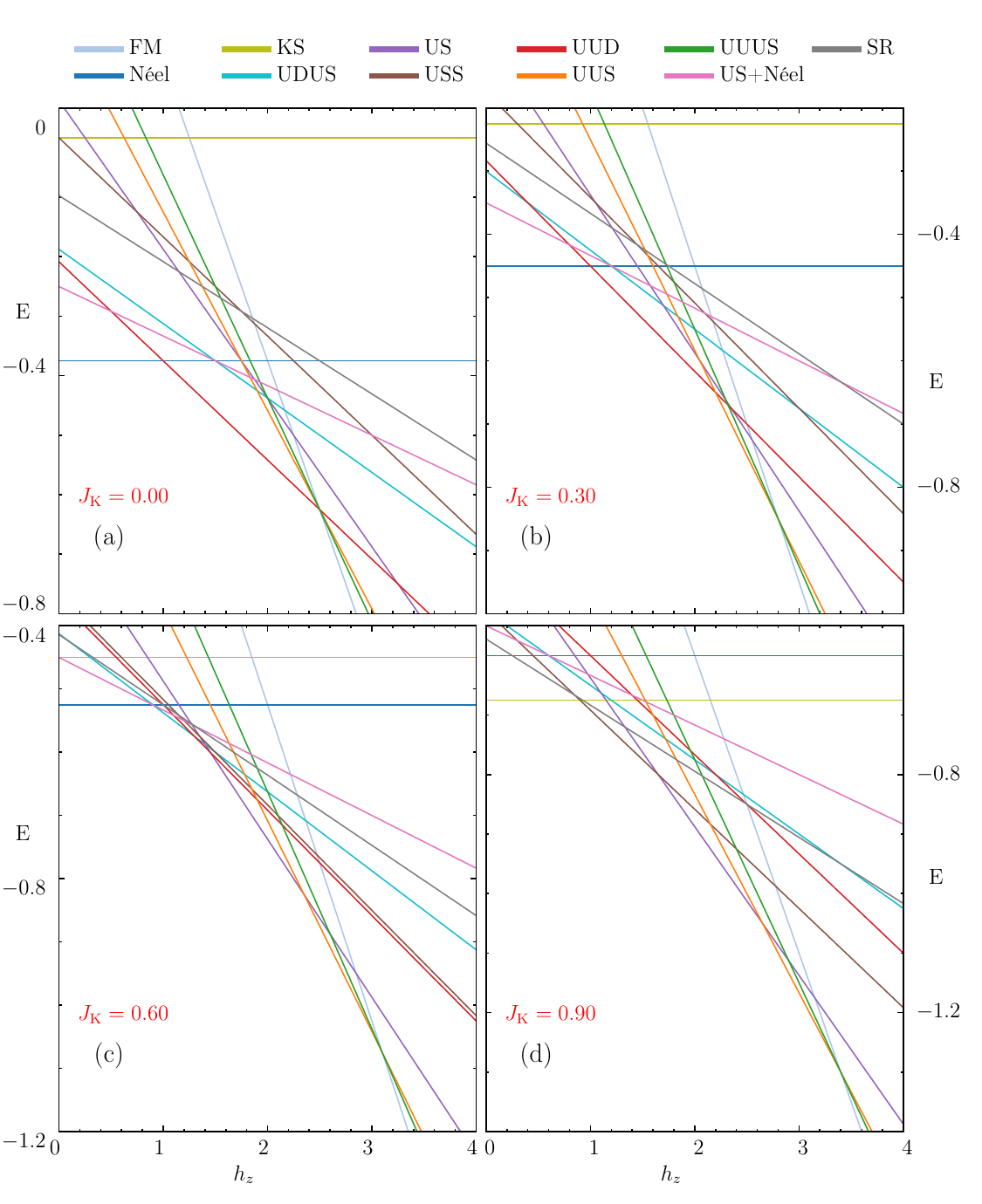}
 \caption{ Variation of total energy of all the candidate variational states with temperature for (a) $J_{\rm K} = 0.00$, (b) $J_{\rm K}=0.30$, (c) $J_{\rm K}=0.60$ and (d) $J_{\rm K} = 0.90$}
\label{fig:S1}   
\end{figure}

\begin{figure}[h!]
\includegraphics[width=0.95\columnwidth,angle=0,clip=true]{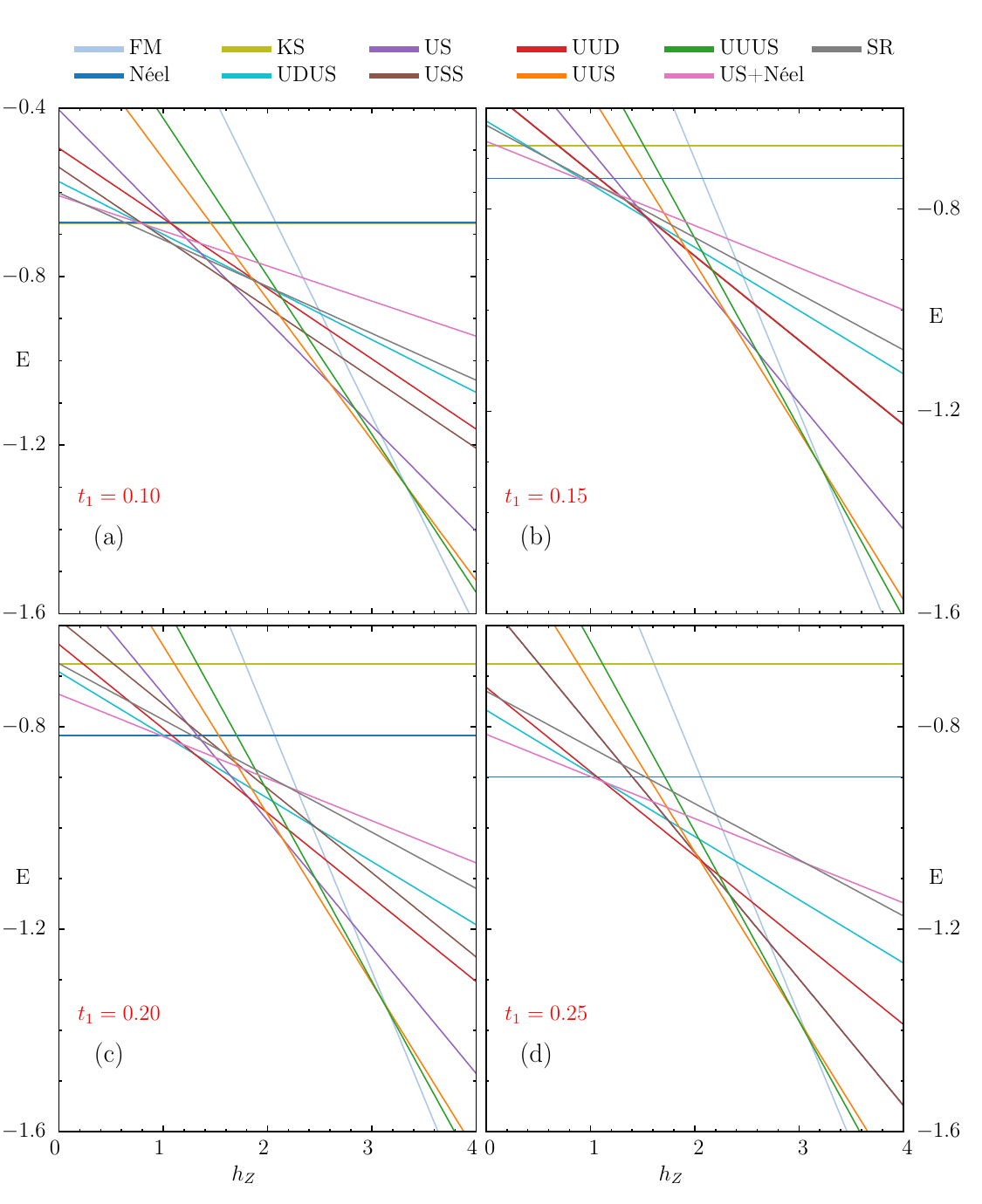}
 \caption{ Variation of total energy of all the candidate variational states with temperature for $J_{\rm K} = 0.90$ , (a) $t_1 = 0.10$, (b) $t_1=0.15$, (c) $t_1=0.20$ and (d) $t_1 = 0.25$}
\label{fig:S2}   
\end{figure}

We first focus on the dependence of the ground state on the magnetic field \(h_z\) for different values of the Kondo coupling strength \(J_{\rm K}\). In Fig.~\ref{fig:S1}(a), we show the energy variation of all the candidate states as a function of the magnetic field \(h_z\) for \(J_{\rm K}=0\). We find that at low fields the N\'eel state is favored, while at intermediate fields the UUD state has the lowest energy, and at higher fields all spins align ferromagnetically. Similar comparisons are also shown for \(J_{\rm K}=0.30\), \(0.60\) and \(90\). To map out the phase diagram in the \(J_{\rm K}\)--\(h_z\) plane, we followed the same process for additional values of \(J_{\rm K}\), with improved accuracy compared to smaller-system simulations.

A similar protocol is followed to investigate the role of finite hopping. In this case, we explore the $t_1$--$h_z$ parameter space by evaluating the energies of all candidate states obtained from the HMC. Note that during all the calculations, $t_2/t_1$ ratio is kept fixed as 1. The energy comparision of all the candidate states for $t_1 =$ \(0.10\), \(0.15\), \(0.20\) and \(0.25\) are shown in Fig. \ref{fig:S2}(a)-(d).  By systematically scanning over a dense value of $t_1$, we identify the energetically favored phases and determine the corresponding phase boundaries.

Overall, the variational calculations serve as an independent and complementary approach to the Monte Carlo simulations, enabling a consistent determination of the ground-state properties across a wide range of parameters while minimizing finite-size artifacts.

\section{Phase Diagrams and Real Space Configuration}
In the main text, we presented the phase diagram of the Kondo necklace model at the maximally frustrated point, \(J_2/J_1 = 1.0\). Here, using variational calculations with all the candidate states obtained from the MC simulations, we show the phase diagram for \(J_2/J_1 = 0.75\) in Fig.~\ref{fig:S3}. We find that all the phases present for \(J_2/J_1 = 1.0\), except the US+N\'eel phase, remain stable in this case as well. However, the phase boundaries of the different phases are shifted. 

Similarly, we followed the same protocol for \(J_2/J_1 = 1.50\). In this case, compared to \(J_2/J_1 = 1.0\), the UDUS phase is absent (see Fig.~\ref{fig:S4}), and the phase boundaries are again shifted. In addition, another phase, namely the up-down-singlet (UDS) phase with \(m=0\) and \(n_{\rm K} = 1/3\), is also found, as shown in Fig.~\ref{fig:S5}.

\begin{figure}[h!]
\centering
\includegraphics[width=0.70 \columnwidth,angle=0,clip=true]{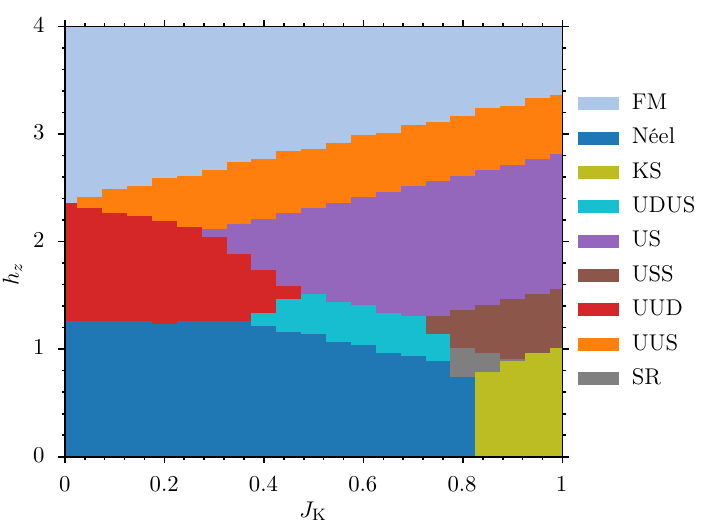}
 \caption{phase diagram in $J_{\rm K}$ - $h_z$ plane for $J_2/J_1 = 0.75$}
\label{fig:S3}   
\end{figure}

\begin{figure}
\centering
\includegraphics[width=0.70 \columnwidth,angle=0,clip=true]{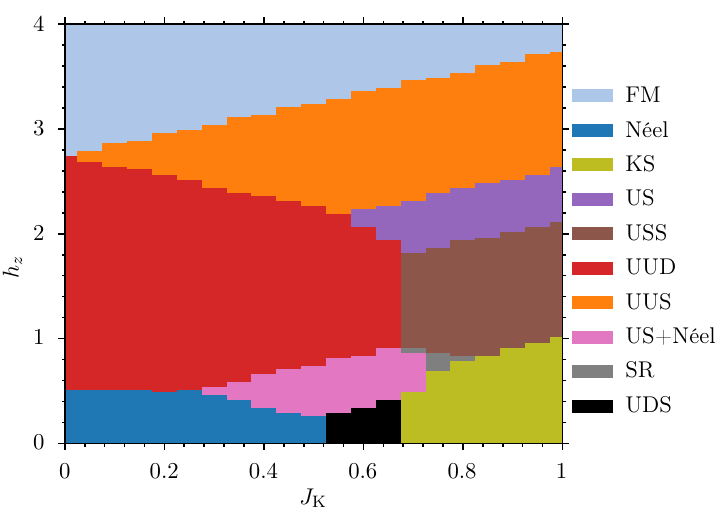}
 \caption{phase diagram in $J_{\rm K}$ - $h_z$ plane for $J_2/J_1 = 1.50$}
\label{fig:S4}   
\end{figure}

\begin{figure}
\centering
\includegraphics[width=0.45 \columnwidth,angle=0,clip=true]{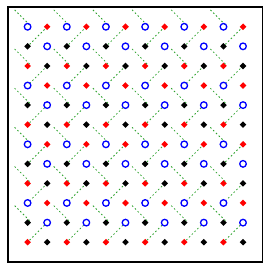}
 \caption{Real-space patterns of spins (red: ${\cal S} = \tfrac{1}{2}$, black: ${\cal S} = -\tfrac{1}{2}$) and singlets (blue circles) of a $12 \times 12$ lattice section corresponding to UDS: $m= 0$, $n_{\rm K} = 1/3$} 
\label{fig:S5}   
\end{figure}

\end{widetext}
\clearpage

\bibliography{merged}



\end{document}